\documentclass[letter]{aa}
\usepackage{siunitx}
\usepackage{float}
\usepackage{adjustbox}
\usepackage{adjustbox}
\usepackage{graphicx}	
\usepackage{amsmath}	
\usepackage{amssymb}
\usepackage{hyperref}
\usepackage{tabularx}
\usepackage{mathptmx}
\usepackage{txfonts}
\usepackage{academicons}
\usepackage[T1]{fontenc}
\usepackage{longtable}
\usepackage{subcaption}
\usepackage[symbol]{footmisc}
\usepackage{multicol, blindtext}
\usepackage{subcaption}
\usepackage[font=small]{caption}

\providecommand{\arcsec}{$^{\prime \prime}$}

\DeclareRobustCommand{\VAN}[3]{#2}
\let\VANthebibliography\thebibliography
\def\thebibliography{\DeclareRobustCommand{\VAN}[3]{##3}\VANthebibliography}
\usepackage{hyperref}
\usepackage{academicons}
\usepackage{xcolor}

\usepackage{calrsfs}
\newcommand{\orcid}[1]{\href{https://orcid.org/#1}{\includegraphics[width=8pt]{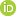}}}

\usepackage{graphicx}
\usepackage{txfonts}
\usepackage{orcidlink}

\begin{document}


      \title{JWST-ALMA Study of a Hub-Filament System in the Nascent Phase}
      
   
  \author{N. K. Bhadari \orcid{0000-0001-8812-8460}
          \inst{1}
                        L. K. Dewangan \orcid{0000-0001-6725-0483}
            \inst{1}
                        O. R. Jadhav \orcid{0009-0001-2896-1896}
            \inst{1,2}
                        Ariful Hoque \orcid{0009-0003-6633-525X}
            \inst{3}
                        L. E. Pirogov \orcid{0000-0001-6251-2421}
            \inst{4}
                        Paul F. Goldsmith \orcid{0000-0002-6622-8396}
            \inst{5}
                        A. K. Maity \orcid{0000-0002-7367-9355}
            \inst{1,2}
                        Saurabh Sharma \orcid{0000-0001-5731-3057}
            \inst{6}
                         A. Haj Ismail \orcid{0000-0003-4941-5154}
            \inst{7}
            \and
                         Tapas Baug \orcid{0000-0003-0295-6586}
            \inst{3}
            }

   \institute{Astronomy $\&$ Astrophysics Division, Physical Research Laboratory, Ahmedabad 380009, India\\
              \email{bhadrinaval@gmail.com}
          \and
              Indian Institute of Technology, Gandhinagar 382355, India
        \and
            Satyendra Nath Bose National Centre for Basic Sciences, Block-JD, Sector-III, Salt Lake, Kolkata-700 106, India
        \and
            Institute of Applied Physics of the Russian Academy of Sciences, 46 Ul’yanov Str., 603950 Nizhny Novgorod, Russia
        \and
            Jet Propulsion Laboratory, California Institute of Technology, 4800 Oak Grove Drive, Pasadena, CA 91109, USA
        \and
            Aryabhatta Research Institute of Observational sciencES (ARIES), Manora Peak, Nainital 263001, India
        \and
            College of Humanities and Sciences, Ajman University, Ajman P.O. Box 346, United Arab Emirates     
           }
            
    \titlerunning{A primordial HFS detected by JWST and ALMA}
    \authorrunning{N. K. Bhadari et al.,}

   \date{Received aaa; accepted bbb}

 
\abstract
{Star clusters, including high-mass stars, form within hub-filament systems (HFSs). Observations of HFSs that remain unaffected by feedback from embedded stars are rare yet crucial for understanding the mass inflow process in high-mass star formation. Using the JWST NIRCAM images, \citet{Dewangan2024} reported that the high-mass protostar G11P1 is embedded in a candidate HFS (G11P1-HFS; $<0.6$ pc).}
{Utilizing ALMA N$_{2}$H$^{+}$(1--0) data, we confirm the presence of G11P1-HFS and study the dense gas kinematics.}
{We analyzed the position-position-velocity ({\it PPV}) map and estimated on-sky velocity gradient ($V_g$) and gravity ($\mathcal{F}_{g}$) vectors. The spatial distribution of gas velocity and H$_2$ column density was examined.
}
{The steep $V_g$ of 5 km s$^{-1}$ pc$^{-1}$ and $-$7 km s$^{-1}$ pc$^{-1}$ toward either side of G11P1-hub, and the decreasing $V_g$ toward the hub, identify G11P1-HFS as a small-scale HFS in its nascent phase. $V_g$ and $\mathcal{F}_{g}$ align along the filaments, indicating gravity-driven flows.}
{This work highlights the wiggled, funnel-shaped morphology of a HFS in {\it PPV} space, suggesting the importance of subfilaments or transverse gas flows in mass transportation to the hub.}
{}


\maketitle
%
\section{Introduction}
\label{sec:intro}

Molecular clouds possess intricate filamentary features, which often intersect, forming web-like structures and creating junctions known as hubs \citep{Myers2009}.
Such structures, named hub-filament systems (HFSs), are widely recognized as potential nurseries of star clusters hosting high-mass stars \citep[$>8M_{\odot}$;][]{Morales2019,Dewangan2020,Kumar2020}.
While several studies support that hubs are fed by matter funneled through filaments \citep[e.g.,][]{Peretto2014,Gomez2014,Yang2023}, the physical mechanisms driving mass assembly within hubs remain poorly understood. This is observationally challenged by the scarcity of pristine HFSs that remain unaffected by feedback from young high-mass stars.
The evidence of widespread existence of HFSs at scales ranging from $\sim$0.1 pc to $>$10 pc signifies that self-similar HFSs exist in multi-scale hierarchy \citep[e.g.,][]{Bhadari2022,Zhou2022,Zhou2023}.
This suggests multi-scale accretion processes in massive star formation \citep[MSF;][and references therein]{Peretto2013,Dewangan2021_w42,Semadeni2024}.
These processes can be divided into two broad categories based on how stars accrete mass from their parent structures: clump-fed and core-fed accretion scenarios. Clump-fed accretion focuses on mass assembly through global clump infall or coherent gas flows, including processes such as competitive accretion \citep[CA;][]{Bonnell2006}, global hierarchical collapse \citep[GHC;][]{Semadeni2019}, and inertial inflow \citep[I2;][]{Padoan2020}. In contrast, core-fed accretion primarily relies on the collapse of unfragmented high-mass starless cores \citep[turbulent core accretion;][]{McKee2003}.
While most observations support clump-fed scenarios, the role of turbulence and gravity, including the physical scales at which they influence the formation of dense structures leading to MSF, remains debatable \citep{Motte2018,Hacar2023}.

\begin{figure*}[htb!]

        \centering
            \includegraphics[width=1\textwidth]{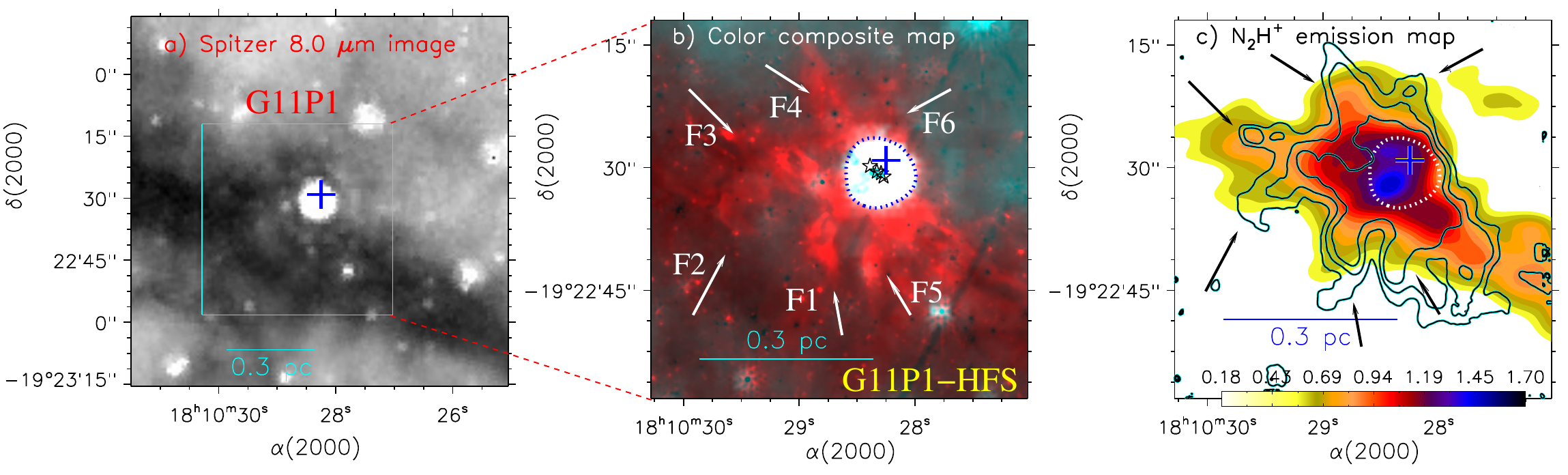}
        
        \caption{Morphology of G11P1-HFS seen in the Spitzer, JWST, and ALMA observations. a) Spitzer 8 $\mu$m image.
        b) Two-color composite map made of the JWST ratio map (F444W/F356W; inverted scale in {\it red}) and Spitzer 8 $\mu$m ({\it cyan}) emission. Stars show the positions of compact radio continuum sources from \citet{Rosero2014}. 
        c) ALMA N$_{2}$H$^{+}$(1--0) integrated emission map overlaid with the footprints of structures seen in the JWST ratio map (see text for details). The color scale bar has a unit of Jy beam$^{-1}$ km s$^{-1}$.
        In each panel, a cross represents the position of the 6.7-GHz methanol maser. A dotted contour in panels ``b'' and ``c'' mark the Spitzer 8 $\mu$m emission feature.
        In panels ``b'' and ``c'', arrows highlight possible filamentary features (F1--F6).
        Scale bar represents a physical scale of 0.3 pc at a distance of Galactic Snake \citep[2.92 kpc;][]{Dewangan2024}).
}
        \label{fig1}

\end{figure*}

Galactic `Snake’ (or IRDC G11.11--0.12) is one of the first discovered infrared-dark clouds \citep{Carey1998} where multiple HFSs at various physical scales are observed \citep[][see also,\citet{Wang2014}]{Dewangan2024}.
Recently, utilising the James Webb Space Telescope (JWST) NIRCAM images, \citet[][hereafter, Paper~I]{Dewangan2024} reported a potential small-scale infrared-dark HFS (G11P1-HFS; extent $\sim$0.55 pc) harbouring a high-mass protostar G11P1 in the Galactic Snake.
They also demonstrated that G11P1 is located within a large-scale HFS (HFS3; extent $\sim$3--5 pc), coinciding with the interaction zone of two cloud components ($V_{\rm LSR} \sim$29 and 31 km s$^{-1}$) that form the Snake\footnote{We refer to large-scale HFSs, those with hub sizes of $\sim$1 pc or more, primarily observed in the Herschel maps \citep[e.g.,][]{Kumar2020}. In contrast, small-scale HFSs, observed with ALMA, exhibit clumps or hubs splitting into elongated structures at sub-parsec scales, resembling HFSs, as noted by \citet{Zhou2022} in the ATOMS survey and \citet{Hacar2018} in the `Orion Integral Filament.'}.
A most massive source ($\sim$254 $M_{\odot}$) identified along Snake's spine using the H$_{2}$ column density map in Paper I (see Fig.7 therein) hosts G11P1 and meets the MSF criteria discussed by \citet{Kauffmann2010}.
Fig.~\ref{fig1}a and \ref{fig1}b present the near-infrared (NIR) view of the physical environment around G11P1 using the Spitzer and JWST images, respectively.

This paper, utilising the Atacama Large Millimeter/submillimeter Array (ALMA) N$_{2}$H$^{+}$(1--0) data, confirms the direct observational evidence of a small scale HFS, G11P1-HFS and improves our understanding of mass assembly in the hubs. 

\section{Data Sets}
\label{data:g11}

This work primarily exploits the N$_{2}$H$^{+}$(1--0) line data from the ALMA Band-3 observations of IRDC\_G011.11-4 (central coordinates: RA= 18h 10m 28.3s, DEC= $-$19d 22m 31s) obtained under the Project Code: 2018.1.00424.S (PI: Gieser, Caroline).
We retrieved the processed data from ALMA archive\footnote{https://jvo.nao.ac.jp/portal/alma/archive.do}.
The angular resolution of the data cube was 4$''$.712 $\times$ 2$''$.588 (0.07 $\times$ 0.037 pc$^{2}$ at $d=2.92$ kpc) with pixel scale of 0$''$.46 \citep[see observation details in][]{Gieser2023}.
To complement the line data, we also utilize the Spitzer 8 $\mu$m image (resolution $\sim$2$''$; or $\sim$5840 au) and the JWST NIRCAM images at 3.563 and 4.421 $\mu$m (resolution $\sim$0$''$.17; or $\sim$500 au) from Paper~I.
Since N$_{2}$H$^{+}$ has high critical density (4.4 $\times$ 10$^{5}$ cm$^{-3}$ at 40 K), it can probe the high density regions \citep[e.g.,][]{Pirogov2003}.
Though consisting of seven hyperfine observationally resolved components \citep[e.g.,][]{Caselli1995,Daniel2006}, N$_{2}$H$^{+}$(1--0) spectra show only three lines in our target region, two of which are triplets (see Fig.~\ref{figA5}).

\section{Analysis and Results}
\label{results:g11}

Fig.~\ref{fig1}b shows the two-color composite image, {\it red:} the JWST ratio map (F444W/F356W), and {\it cyan:} the Spitzer 8 $\mu$m emission map of the G11P1 environment.
The Spitzer/JWST ratio map (inverted scale) exhibits a strong morphological correlation with the mid-infrared emission, tracing warm dusty features \citep[see Fig.4 in][and \citet{Dewangan2023}]{Bhadari2020}.
At least six elongated structures (F1--F6; filaments, hereafter) are evident that connect to the central region hosting several compact radio continuum sources (CRCSs) including G11P1 and a 6.7 GHz methanol maser emission \citep[see also][]{Rosero2014}. 
This traces a potential small-scale HFS ($\sim$0.55 pc; Paper I), which needs confirmation through molecular line observations.

\subsection{ALMA N$_{2}$H$^{+}$(1--0) emission}
\label{n2h_emission}
To compare the morphology of structures seen in the JWST images with the dense gas structures, we present the overlay of the JWST ratio map contours on the N$_{2}$H$^{+}$(1--0) integrated emission map in Fig.~\ref{fig1}c. The N$_{2}$H$^{+}$(1--0) emission is integrated over velocity range from 25 to 32 km s$^{-1}$ considering the major peak of N$_{2}$H$^{+}$(1--0) triplets (see Sec.~\ref{data:g11}).
The N$_{2}$H$^{+}$(1--0) spectrum averaged over the HFS is shown in Fig.~\ref{figA5}.
To reduce noise and enhance the structures visible in the JWST images, we applied median filtering with a 6-pixel width, followed by smoothing the resulting map with a 3 $\times$ 3 pixel$^{2}$ box-car kernel. These parameters were chosen to achieve optimal image presented in Fig.~\ref{fig1}b (see also contours in Fig.~\ref{fig1}c).
We note that there is a significant similarity between the morphology of N$_{2}$H$^{+}$ emission and the structures traced by the JWST. This directly confirms the G11P1-HFS as a HFS seen at small scale ($<0.6$ pc). The N$_{2}$H$^{+}$(1--0) moment maps (e.g., integrated intensity, intensity weighted centroid velocity ($V_{\rm LSR}$), intensity weighted linewidth, i.e., full width at half maximum (FWHM; $\Delta V_{\rm LSR}$) maps) are presented in Figs.~\ref{figA1}a--\ref{figA1}c, respectively. 
For comparison, we also present the moment-2 map derived using the N$_{2}$H$^{+}$(1--0) singlet F$_{\rm 1}$ F = 01--12 ($V_{\rm LSR}=$[18, 23] km s$^{-1}$) in Fig.~\ref{figA1}d.

\subsubsection{Position-Position-Velocity ({\it PPV}) Map}
\label{g11p1_hub}

\begin{figure}[htb!]
    \centering
\includegraphics[width=0.5\textwidth]{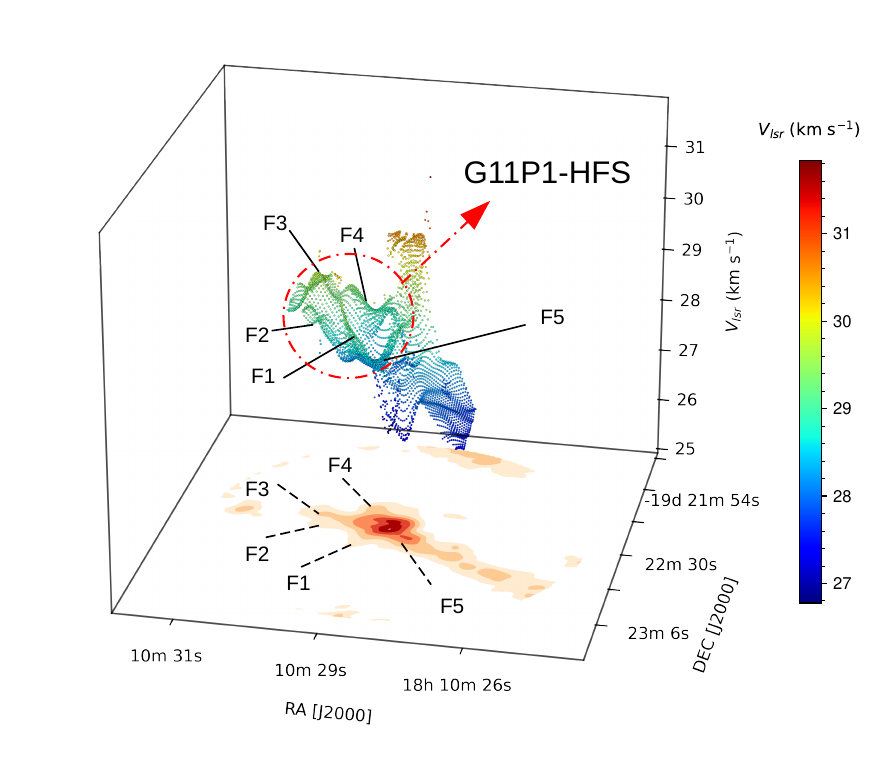}
    \caption {The {\it PPV} map of ALMA N$_{2}$H$^{+}$(1--0) toward G11P1-HFS derived using the \texttt{SCOUSEPY}. The {\it PP} space at the bottom of the map displays the integrated intensity map for the entire structure. Five filaments comprising G11P1-HFS are marked. 
    The 3D view of the {\it PPV} map is available \href{https://drive.google.com/file/d/1OgHx8cJIRgpyi5hrC2cGOg7sb6nTUCQQ/view?usp=sharing}{here}.
    }
    \label{fig2}
\end{figure}

To disentangle the multiple velocity components, we performed the spectral decomposition of the N$_{2}$H$^{+}$(1--0) emission using \texttt{SCOUSEPY}\footnote{https://scousepy.readthedocs.io/en/latest/} tool \citep[][]{Henshaw2016,Henshaw2019}.
The fitted centroid velocity in each pixel is used to generate a {\it PPV} map, which is shown in Fig.~\ref{fig2}. 
The five potential hub-forming filaments (F1--F5) are marked in the {\it PPV} map and in the moment-0 map shown at the bottom face of {\it PPV} cube. A candidate filament, F6 observed in the JWST (Fig.~\ref{fig1}b) may be blended with a potential outflow feature (see gas structure in $V_{\rm LSR}$ $>$31 and $<$28 km s$^{-1}$), so we excluded it from further analysis.
Additionally, the F5 feature is not well resolved in the ALMA N$_{2}$H$^{+}$(1--0) map, therefore, a tentative rectangular box is used for its analysis (see Sec.~\ref{gas_kin}).
Signature of velocity variations is evident along the filaments joining hub.

\begin{figure}[htb!]
    \centering
	\includegraphics[width=0.5\textwidth]{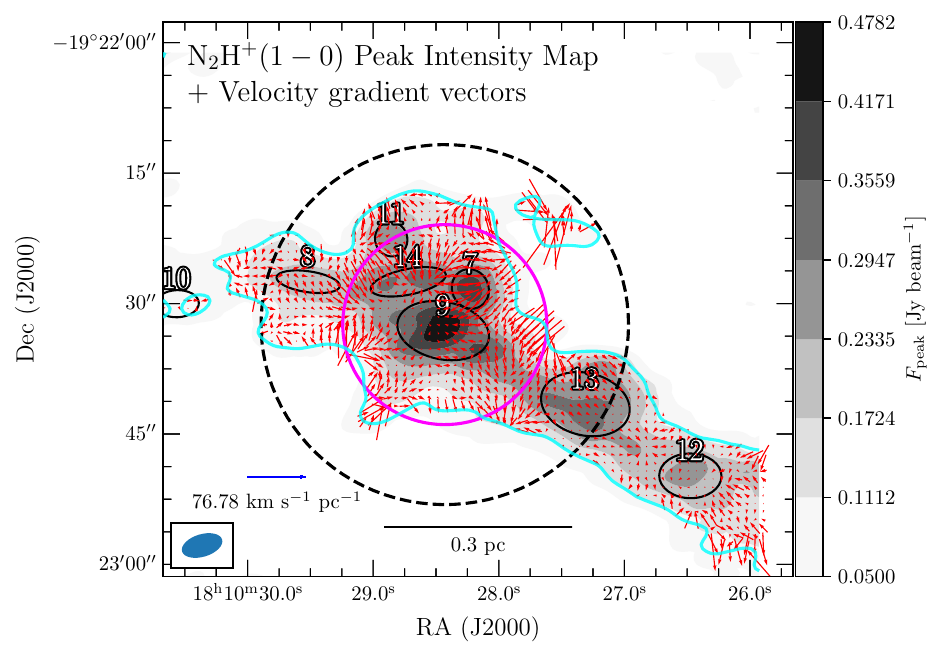}
    \caption {Overlay of velocity gradient ($V_g$) vectors, and \texttt{dendrogram} leaves on the N$_{2}$H$^{+}$(1--0) peak intensity map. The arrow-heads point to local blueshifted velocity material. The reference $V_g$ vector and ALMA beam are shown in the lower left corner. Cyan contour presents the N$_{2}$H$^{+}$ emission at level of 0.2 Jy beam$^{-1}$ km s$^{-1}$. 
    Two circular regions (radii $\sim$12$''$ ($\sim$0.17 pc) and 22$''$ ($\sim$0.3 pc)) typically mark the boundaries of hub and G11P1-HFS (see text and Fig.~\ref{fig4}).
}
    \label{fig3}
\end{figure}

To quantify the on-sky velocity variation, we first derived the magnitude of velocity gradient ($V_g$) along RA-($\mathcal G_\alpha$) and DEC-direction ($\mathcal G_\delta$) as the second order central differences of $V_{\rm LSR}$.
We estimated the resultant magnitude and position angle as $V_{g}=\sqrt{\mathcal G_\alpha^2+\mathcal G_\delta^2}$ and $\theta V_{g}= \frac{180\degr}{\pi}\times {\rm arctan}\left(\frac{\mathcal G_\delta}{\mathcal G_\alpha}\right)$, respectively \citep[see also][]{Eswaraiah2020}.
Fig.~\ref{fig3} displays the overlay of velocity gradient vectors\footnote[1]{The velocity gradient vectors are intentionally drawn pointing to decreasing velocity to make easy comparison with the {\it PPV} map shown in Fig.~\ref{fig2}.} with 2 $\times$ 2 pixel$^{2}$ averaging on the peak-intensity map. 
The vectors converge to at least two peak intensity zones in the G11P1-hub, where \texttt{dendrogram} leaves (or cores) \#L9 ($\sim$22 $M_\odot$) and \#L14 ($\sim$8 $M_\odot$) are also identified (see following text).
A noticeable feature is the gradual decrease of vector lengths as one moves toward these cores (or G11P1-hub centers). {This feature is further quantified in Sec.~\ref{gas_kin} and Fig.~\ref{fig4}}.
We utilized the \texttt{python} based \texttt{astrodendro}\footnote{https://dendrograms.readthedocs.io/en/stable/} package to identify the hierarchical substructures in N$_{2}$H$^{+}$(1--0)  {\it PPV} space. 
The input parameters, {\it min\_value}, {\it min\_delta}, and {\it min\_npix} were set to be 5-sigma, 1-sigma, and 65 pixels, respectively, where 1-sigma ($\sim$14 mJy beam$^{-1}$) is the rms of the data.
The dendrogram leaves/cores are overlaid in Fig.~\ref{fig3}, with their physical parameters listed in Table~\ref{tab:cores}. The \texttt{dendrogram} tree of hierarchical structures (i.e., leaves and branches) is shown in Fig.~\ref{figA3}. 

\subsection{Gas-kinematics of G11P1-HFS}
\label{gas_kin}

We derived the N$_{2}$H$^{+}$ column density ($N(\rm N_{2}H^{+})$) and excitation temperature maps ($T_{\rm ex}$) by modeling the fine structure components of N$_{2}$H$^{+}$(1--0) emission with \texttt{XCLASS} \citep{Moller2017}.
The conversion factor, $\mathcal{X}$= 3$\times 10^{-10}$ was used to derive the H$_2$ column density ($N\rm (H_{2})$) from $N(\rm N_{2}H^{+})$ \citep{Caselli2002}. Figs.~\ref{figA2}a and ~\ref{figA2}b present the $N\rm (H_{2})$ and $T_{\rm ex}$ maps, respectively. 
Following the column density-mass relation \citep[see Eq.1 in][]{Bhadari2020}, the mass of the elongated feature enclosed by the $N\rm (H_{2})$ level of $3\times10^{22}$ cm$^{-2}$ is estimated to be $\sim223~M_{\odot}$, which agrees with the mass of the dense source hosting G11P1 (see Sec.\ref{sec:intro}).
The derived $N(\rm H_{2})$ and mass may have uncertainties of about 60-70\%, due to potentially lower N$_{2}$H$^{+}$ abundance from chemical differentiation \citep{Zinchenko2009} and high uncertainties itself in the calculation of $\mathcal{X}$ for both low-mass and high-mass core samples \citep{Caselli2002,Pirogov2013}.

\begin{figure*}[htb!]
\begin{subfigure}{0.5\linewidth}
  \centering
\includegraphics[width=1\textwidth]{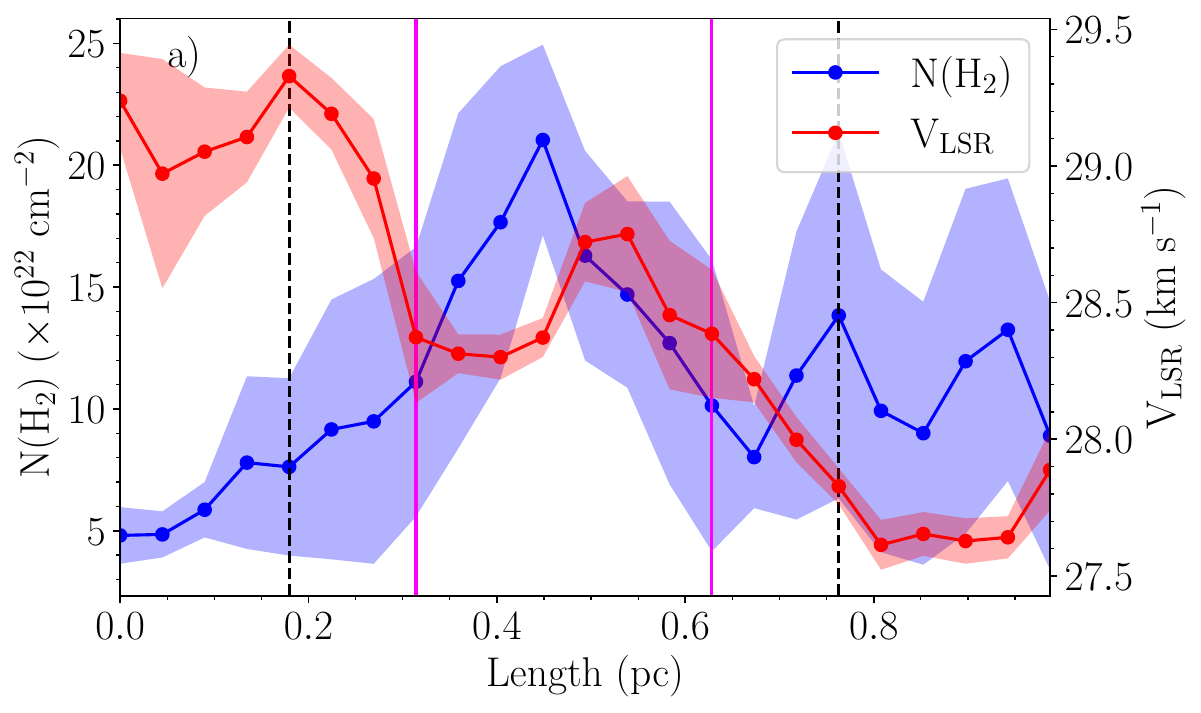}
\end{subfigure}
\begin{subfigure}{0.5\linewidth}
  \centering
\includegraphics[width=1\textwidth]{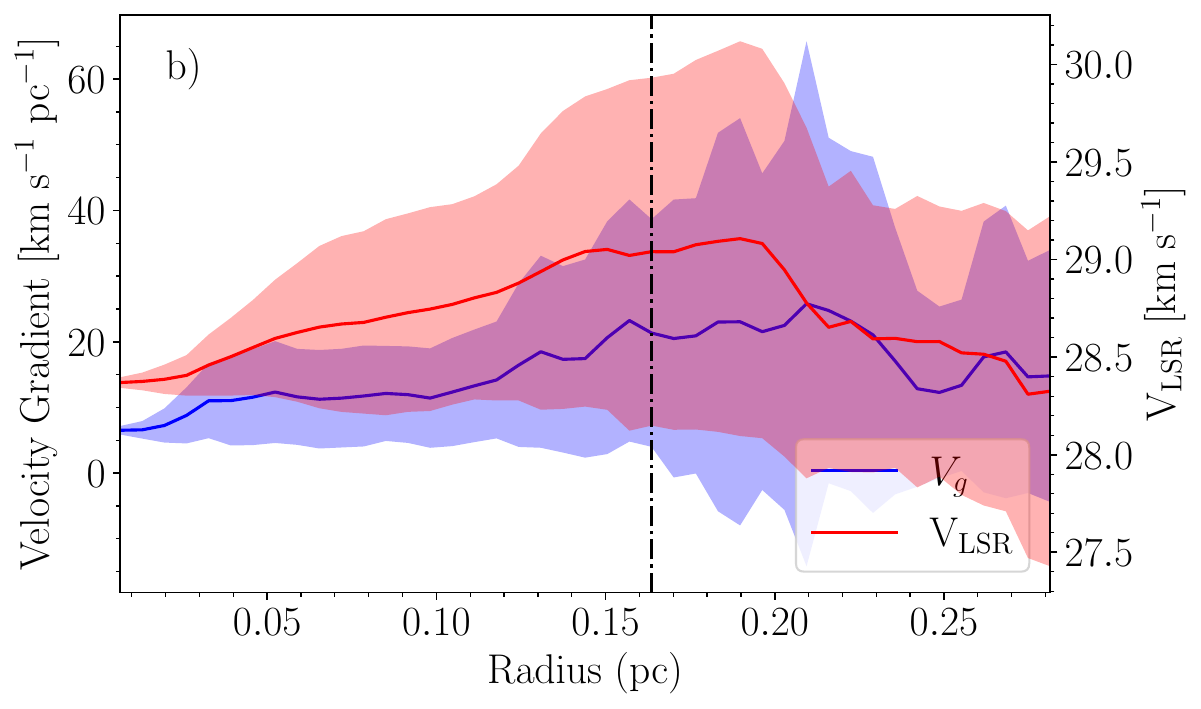}
\end{subfigure}

\caption{a) Distribution of $N(\rm H_{2})$ and $V_{\rm LSR}$ along the rectangular strip shown in Fig.~\ref{figA1}a. 
The dots and shaded regions represent the mean and standard deviation range for each box in the strip.
Dashed and solid vertical lines mark the boundary of two concentric circular regions shown in Fig.~\ref{fig3}.
b) Radial plots for $V_{\rm LSR}$ and $V_{g}$ drawn from the center of concentric circles (Fig.~\ref{fig3}). Vertical dotted-dashed line show the boundary of inner circle in Fig.~\ref{fig3}.
} 
\label{fig4}
\end{figure*}

Fig.~\ref{fig4}a presents the $N\rm (H_{2})$ and $V_{\rm LSR}$ distribution along a rectangular strip (from north-east to south-west direction) marked in Fig.~\ref{figA1}a.
A notable feature includes the simultaneous peak and dip in the $N\rm (H_{2})$ and $V_{\rm LSR}$ profiles at G11P1-HFS center (or \#L9), respectively.
The steep velocity gradient is estimated to be $-7$ km s$^{-1}$ pc$^{-1}$ and $4.7$ km s$^{-1}$ pc$^{-1}$ toward north-east and south-west side of the G11P1-HFS center, respectively.
This strongly favour the presence of a mass accreting object \citep[e.g.,][]{Zhou2023}.
We present azimuthally averaged distribution of $V_{\rm LSR}$ and $V_{g}$ toward G11P1-hub region in Fig.~\ref{fig4}b. 
While the steep slope of $V_{\rm LSR}$ indicates material flow toward hub, the change in $V_{g}$ might indicate nature of converging flows forming a hub (see Sec.~\ref{g11p1:earlyHFS}). 
The $N\rm (H_{2})$ and $V_{\rm LSR}$ profiles for the hub-joining filaments (F1--F5) are shown in Fig.~\ref{figA7}, indicating a significant velocity gradient toward the hub region.
Almost all the $V_{\rm LSR}$ profiles (except F1) show a sharp velocity gradient ($\sim$8--15 km s$^{-1}$ pc$^{-1}$) upto $\sim$0.05 pc from the hub (see the last panel in Fig.~\ref{figA7}). This may be indicative of the hub radius; however, it is important to note that the profiles are not drawn from a common center due to the presence of at least two hub centers.

In order to estimate whether the gas flow along the hub composing filaments is driven by local gravity, we estimated the sky-projected gravitational force vectors ($\mathcal{F}_{g}$) using $N\rm (H_{2})$ map following the approach discussed in \citet[][see Eq.3 therein]{Wang2022}.
The low relative angle between on-sky $V_g$ and $\mathcal{F}_{g}$ vectors along filaments (see Fig.~\ref{figA4}a; and details in Appendix~\ref{gravity:dis}) hints the supportive role of gravity on mass assembly toward G11P1-HFS \citep[e.g.,][]{Wang2022,Maity2024}.

\section{Discussion}
\label{dis:major}

\subsection{Gas inflow along filaments}
Material inflow along filaments is often inferred by the gradual change in gas velocity. Though the knowledge of the actual velocity gradient along filaments is elusive due to the unknown inclination of filaments, the on-sky $V_{\rm LSR}$ variation along filaments is a key proxy used in several studies \citep[e.g.,][]{Kirk2013,Chen2019}.
The longitudinal gas flow along filaments may be driven by either the self-gravity of embedded clumps/cores, pressure-driven flows (or turbulence), the role of magnetic field, or a combination of these factors. At smaller physical scales, the self-gravity of central hubs plays an important role \citep[e.g.,][and references therein]{Zhou2022,Maity2024}.
The signature of the velocity gradient along filaments in G11P1-HFS hints at the mass assembly process. The mass accretion rate along filaments (see boxed regions in Fig.~\ref{figA1}c) was found in the range of 0.92 to 1.44 $M_{\odot}~{\rm Myr^{-1}}$ following the equation $\dot{M}= \frac{\Delta{V_{\rm obs}}M}{tan(\alpha)}$, where $\Delta{V_{\rm obs}}$, $M$, and $\alpha$ are the observed velocity gradient along filament (0.075--0.12 km s$^{-1}$ pc$^{-1}$), filament mass (7.47--18.56 $M_{\odot}$), and inclination angle of filament in the sky plane (assumed to be 45$\degr$), respectively \citep[see details in][]{Kirk2013}.
\citet{Padoan2020} found that the mass-flow rate around prestellar cores increases linearly with size, with an average value of $\sim$10 $M_{\odot}~{\rm Myr^{-1}}$ at 0.1 pc and $\sim$100 $M_{\odot}~{\rm Myr^{-1}}$ at 1 pc. Similar results have also been reported in observational studies \citep[e.g.,][]{Peretto2014,Yuan2018}. Our derived values are two orders of magnitude less than the reported values. This discrepancy may be due to missing flux, which affects the mass and size of the filaments.
Nevertheless, these results align with the GHC and I2 scenarios, supporting the idea that low-mass prestellar cores grow into high-mass protostars through continuous mass accretion from parent clumps or filamentary structures.
Gravity appears to play a dominant role in mass accretion along filaments (see Fig.~\ref{figA4}), which is consistent with both the GHC and I2 models at core scales.
However, to unveil the acting scale of gravity and turbulence, a systematic study consisting of interferometric and single-dish line observations can help \citep[e.g.,][]{Hacar2018}, which is beyond the scope of the current work.

\subsection{G11P1-HFS: A HFS at nascent phase}
\label{g11p1:earlyHFS}

Typically, once a high-mass star forms, it tends to disrupt the parent clump or cloud structure through its energetic feedback processes \citep[e.g., strong ionization through extreme UV photons, stellar winds, and outflows;][]{Motte2018,Schneider2020}, and thereby diminishing the signature of their formation processes.
In this context, tracing the undestroyed HFS hosting young high-mass protostar becomes important to understand the mass assembly in MSF processes.
G11P1-HFS is one such site where the high-mass protostar G11P1 along with cluster members are embedded in a small scale HFS.
It has only been possible by the high-resolution ALMA observations to probe the inner region of the parsec scale hubs \citep{Kumar2020}.
\citet{Zhou2022} identified small-scale HFSs ($\geq0.1$ pc) in a large sample of protoclusters using H$^{13}$CO$^{+}$(1--0) under ATOMS survey. These observations are in line with our outcomes and the discussion of \citet{Kumar2022}, which supports that HFSs are self-similar hierarchical structures observed at a physical scale ranging from protostellar envelope to the clump/cloud scales \citep[i.e., $\geq0.1$ pc;][and references therein]{Zhou2023,Bhadari2022}.
The ``V''--shaped $V_{\rm LSR}$ profile in the direction of the hub, where the $N(\rm H_{2})$ profile peaks (see Fig.\ref{fig4}a), is a key signature of a spherically-symmetric infall motion or mass-accreting hub \citep[][]{Zhou2023}.
Furthermore, the gradual decline of $V_g$ toward dense cores (Figs.~\ref{fig3} and \ref{fig4}b) may indicate the coalescing nature of filaments toward the hub. This can be caused by the shock dissipation and exchange of momentum between filaments with different velocities resulting in low $V_g$ at flow converging zones \citep{Fukui2021}. 
The regularly spaced cores \#L10, \#L8, \#L9, \#L13, and \#L12 along the major filament spine (mean projected separation $\sim$0.22$\pm$0.02 pc; Fig.~\ref{figA1}a) suggest their formation through filament fragmentation processes \citep{Inutsuka1992}. This has also been observed in the vicinity of the high-mass protostar W42-MME \citep{Bhadari2023}, hinting that pristine hubs, and thus high-mass stars, form at the centers of their parent gas streams.
The higher values of $\alpha_{\rm vir}$ (Table~\ref{tab:cores}) suggest that these dendrogram structures may be transient and indicate a dynamic and rapid mass assembly process in the HFS.

\begin{figure}[htb!]
    \centering
	\includegraphics[width=0.5\textwidth]{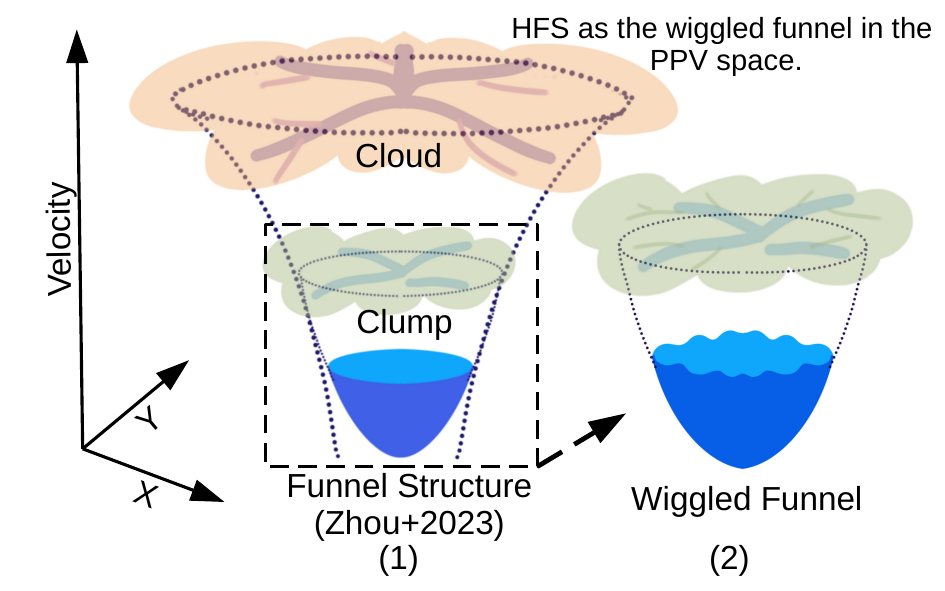}
    \caption {Schematic diagram of hierarchical HFSs in {\it PPV} space as a funnel structure: (1) discussed by \citet{Zhou2023}, (2) wiggled funnel, revealing the role of sub-filaments.
}
    \label{fig5}
\end{figure}

The absence of extended radio emission, the presence of CRCSs, and mass-accreting structures suggest that G11P1-HFS is in its nascent phase.\footnote{We refer to the nascent phase of the HFS, which is at its lower hierarchy and still preserves the signature of multiple accreting filaments.} In comparison, the nearby Mon R2 HFS ($d=830$ pc), which has a similar physical scale ($< 0.8$ pc; \citealt{Morales2019,Kumar2022,Dewangan2024monr2}), appears to be relatively evolved, as it hosts an ultra-compact H{\sc ii} region and an embedded stellar cluster with at least two B-type stars at its hub center.


\subsection{Wiggled funnel: Signature of HFSs in {\it PPV} space}
\label{flower_model}

Regardless of their extents, HFSs are recognized as converging filaments toward the position-position or 2D-maps. However, their significance in mass assembly is underscored when these systems exhibit a ``V'' shape in {\it PV} space or a funnel morphology in {\it PPV} space \citep[see Fig.9 in][]{Zhou2023}. 
The funnel signature may be feeble at cloud scales due to external feedback factors such as turbulence, outflows, and H{\sc ii} region feedback. At core scales, this signature can be discernible during early star-formation stages.
G11P1-HFS presents an ideal site for such observations. 
Diverging from the ideal funnel structure, our observations reveal an azimuthally wiggling funnel structure of G11P1-HFS in {\it PPV} space. This is supported by the study of G333 by \citet[][see Fig. 6]{Zhou2023} that HFSs display wiggling features in {\it PPV} map. This implies the role of sub-filaments that feed the major filament connected to the hub, which are generally observed as transverse structures away from the hub.
Fig.~\ref{fig5} presents the schematic diagram of molecular clouds and hierarchical filamentary substructures as they appear in {\it PPV} space.

\section{Conclusions}
\label{sec:conc}

ALMA N$_{2}$H$^{+}$(1--0) observations confirm the presence of IR-dark G11P1-HFS ($< 0.6$ pc) that was previously identified using the JWST images.
In addition to the ``V''--shaped velocity profile toward the G11P1-hub (i.e., $V_g$ of $-7$ and 5 km s$^{-1}$ pc$^{-1}$), the wiggled funnel feature observed in {\it PPV} space indicates the presence of a mass-accreting hub and role of transverse gas flow.
The low relative angle between on-sky $V_g$ and $\mathcal{F}_{g}$ vectors hints at the gravity driven inflow process along filaments.

\begin{acknowledgements}
We thank the anonymous reviewer for useful comments and suggestions, which improved the quality of manuscript. The research work at the Physical Research Laboratory is funded by the Department of Space, Government of India. 
A portion of this research was carried out at the Jet Propulsion Laboratory, California Institute of Technology, under a contract with the National Aeronautics and Space Administration (80NM0018D0004). L.E.P. acknowledges the support of the Russian Science Foundation (project 24-12-00153).
A.H.I. acknowledges the support from Ajman University, IRG No: [DGSR Ref. 2024-IRG-HBS-7].
This paper makes use of the following ALMA data: ADS/JAO.ALMA\#2018.1.00424.S. ALMA is a partnership of ESO (representing its member states), NSF (USA) and NINS (Japan), together with NRC (Canada) , MOST and ASIAA (Taiwan), and KASI (Republic of Korea), in cooperation with the Republic of Chile. The Joint ALMA Observatory is operated by ESO, AUI/NRAO and NAOJ.
This work is based (in part) on observations made with the NASA/ESA/CSA JWST . The data were obtained
from the Mikulski Archive for Space Telescopes at the Space
Telescope Science Institute, which is operated by the Association of
Universities for Research in Astronomy, Inc., under NASA contract
NAS 5–03127 for JWST . These observations are associated with the
programme \#1182.
This research made use of \texttt{astrodendro}, a \texttt{Python} package to compute dendrograms of Astronomical data (http://www.dendrograms.org/).
This work made use of Astropy:\footnote{http://www.astropy.org} a community-developed core Python package and an ecosystem of tools and resources for astronomy \citep{astropy:2013, astropy:2022}.

\end{acknowledgements}

\bibliographystyle{aa}
\bibliography{references_g11}

\begin{appendix}
\twocolumn

\section{Distribution of on-sky gravity, velocity, and intensity vectors}
\label{gravity:dis}

The relative orientation between possible pairs of $\mathcal{F}_{g}$, $V_g$, and intensity gradient ($I_g$; derived using moment-0 map) was estimated neglecting the true vector direction, thereby converging angle difference ($\Delta\theta_{i,j}$) is in the range of 0--90$\degr$, where $i$ and $j$ refer to two different parameters (i.e., velocity gradient ({\it v}), intensity gradient ({\it i}), and gravity ({\it g})).
Fig.~\ref{figA4}a displays the spatial distribution of $\Delta\theta_{v,g}$ toward G11P1-HFS.
To highlight the filament positions, we have overlaid the JWST ratio map contour (see Fig.~\ref{fig1}c). Interestingly, along the filament spines, the $\Delta\theta_{v,g}$ values are significantly low, indicating that the $V_g$ and $\mathcal{F}_{g}$ vectors are well aligned in these areas compared to other locations.
Fig.~\ref{figA4}b shows the histogram distribution of $\Delta\theta_{v,g}$, $\Delta\theta_{i,g}$, and $\Delta\theta_{i,v}$ toward the entire G11P1-HFS.
The $\Delta\theta_{v,g}$ shown in Fig.~\ref{figA4}b have mean/median values of approximately 41$\degr$/39$\degr$, highlighting the dominant role of gravity in mass accretion process.

\begin{figure}[htb!]
    \centering
	\includegraphics[width=0.5\textwidth]{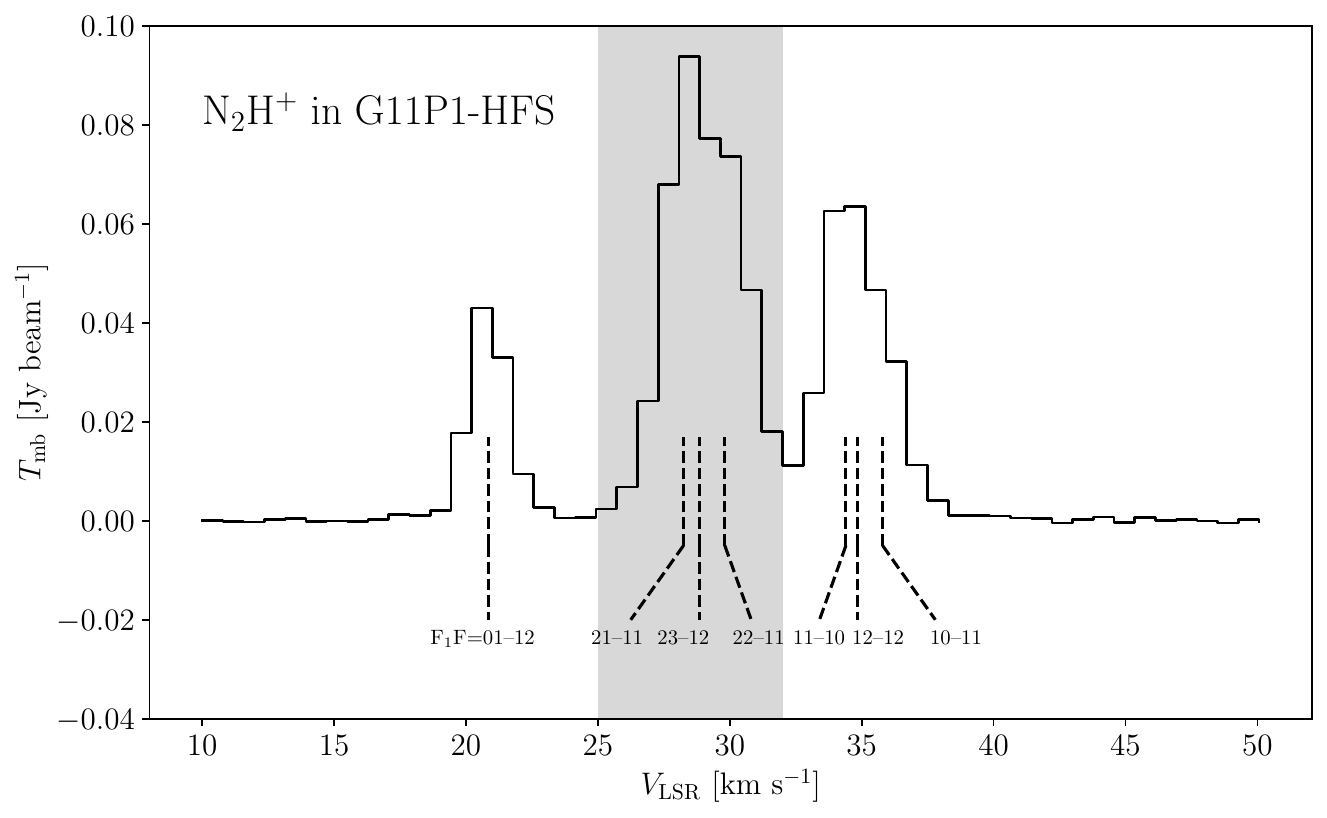}
    \caption {N${2}$H$^{+}$(1--0) spectrum averaged over the G11P1-HFS region (see the larger circle in Fig.~\ref{fig3}). The seven hyperfine components are labeled. The shaded region shows the central peak with $V_{\rm LSR}$ ranging from 25 to 32 km s$^{-1}$, which is used for the gas kinematics study (Sec.~\ref{gas_kin}).
}
    \label{figA5}
\end{figure}

\begin{figure*}[htb!]
\begin{subfigure}{0.5\linewidth}
\includegraphics[width=1\textwidth]{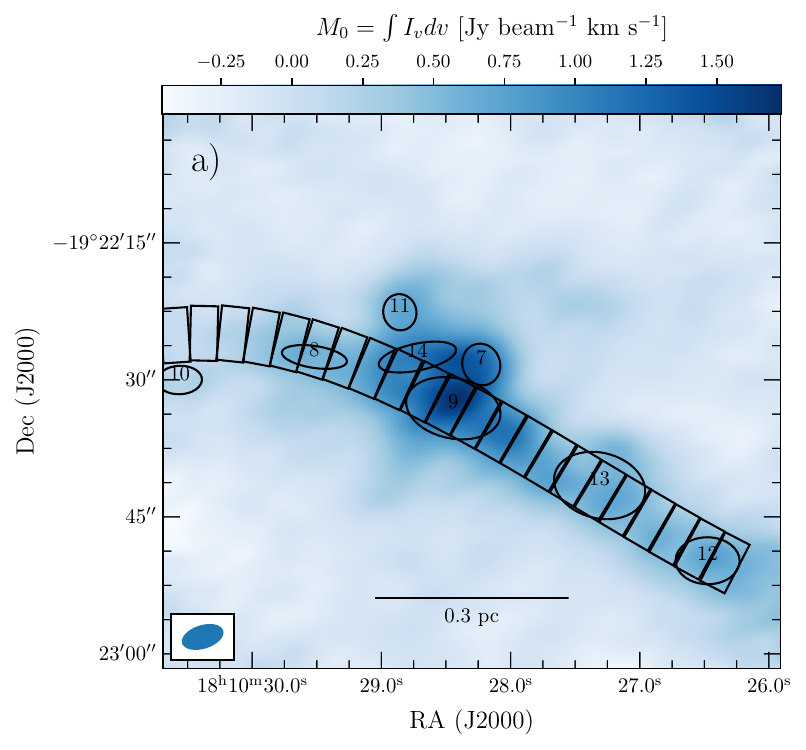}
\end{subfigure}
\begin{subfigure}{0.5\linewidth}
\includegraphics[width=1\textwidth]{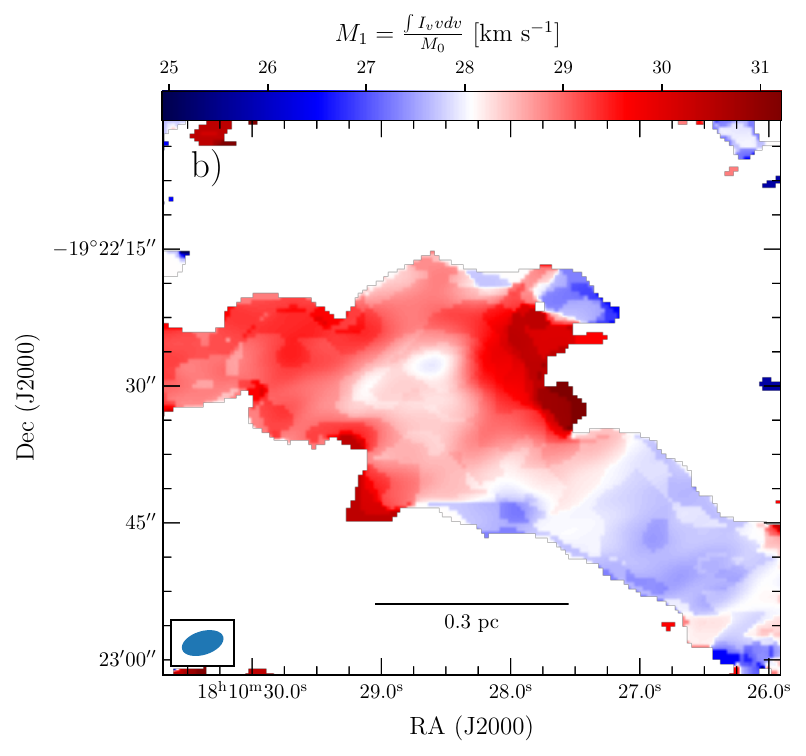}
\end{subfigure}
\begin{subfigure}{0.5\linewidth}
\includegraphics[width=1\textwidth]{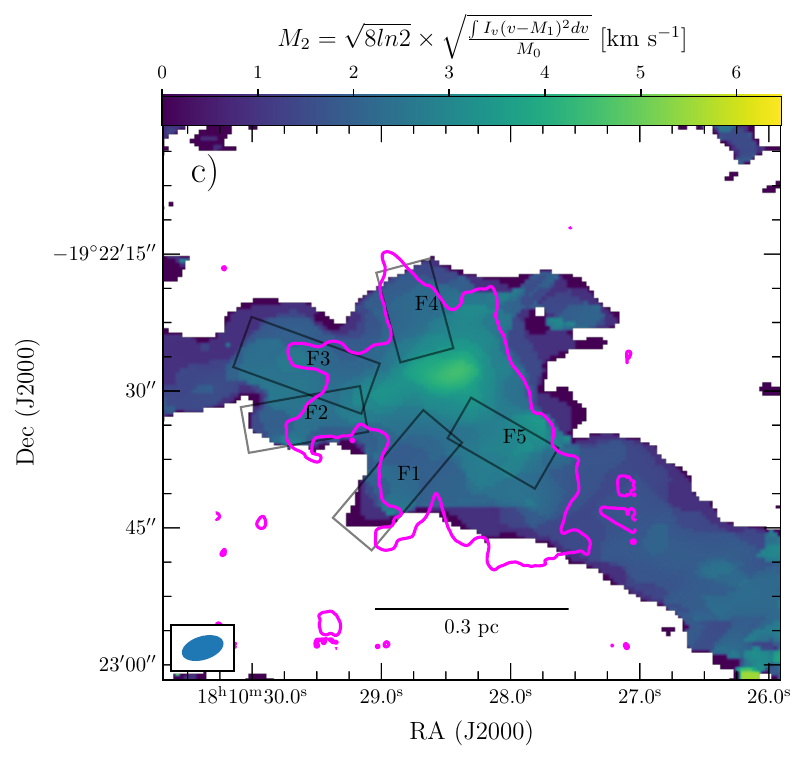}
\end{subfigure}
\begin{subfigure}{0.5\linewidth}
\includegraphics[width=1\textwidth]{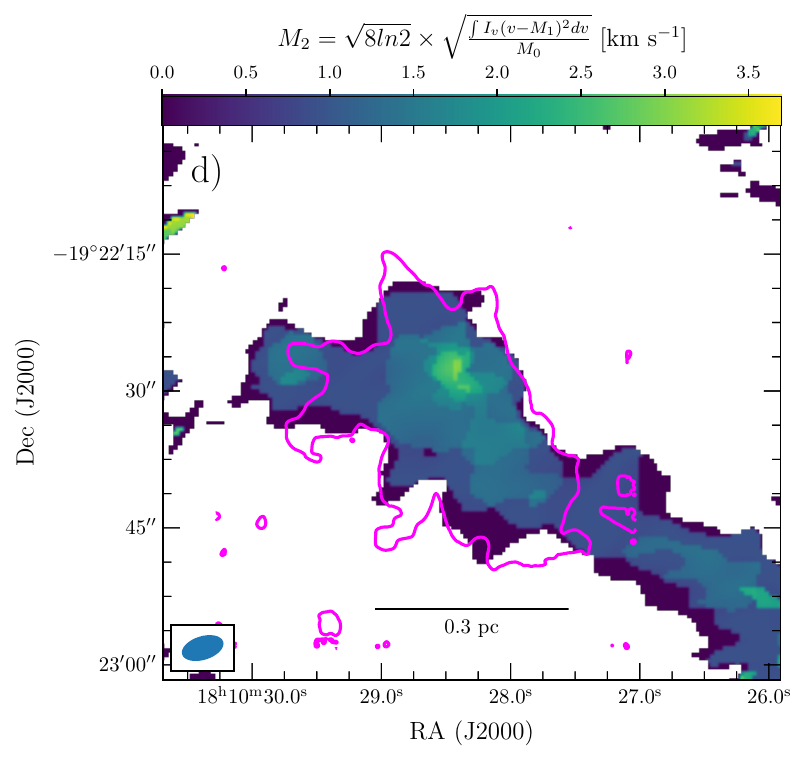}
\end{subfigure}


\caption{(a--c) Moment-0/1/2 maps of N$_{2}$H$^{+}$(1--0) emission derived using central peak (see shaded area in \ref{figA5}). d) Moment-2 map of N$_{2}$H$^{+}$(1--0) emission made using singlet F$_{\rm 1}$ F = 01--12.
The apertures of identified cores, and a rectangular strip are overlaid in panel (a). In panel (c), rectangular boxes mark areas of hub-composing filaments (F1-F5) and a contour displays structure traced by the JWST (see Fig.~\ref{fig1}c).
} 
\label{figA1}
\end{figure*}
%

\begin{figure*}[htb!]
\begin{subfigure}{0.5\linewidth}
  \centering
\includegraphics[width=1\textwidth]{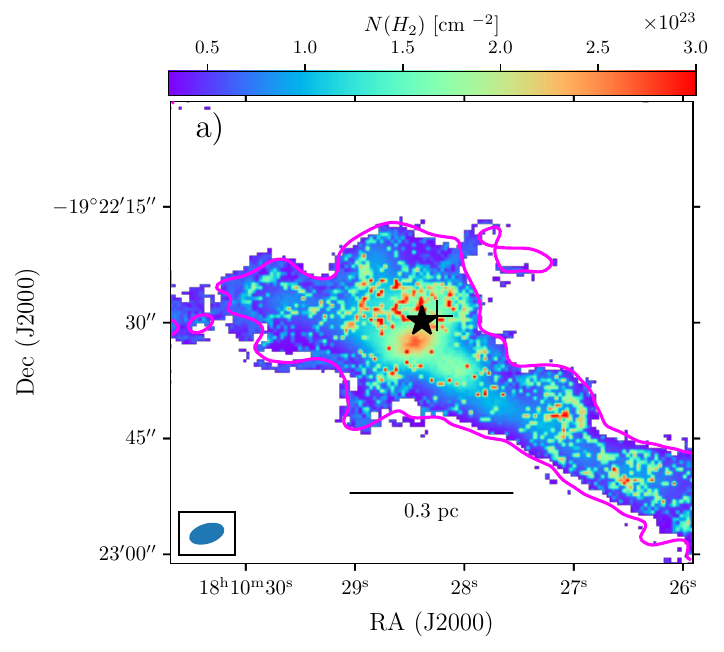}
\end{subfigure}
\begin{subfigure}{0.5\linewidth}
  \centering
\includegraphics[width=1\textwidth]{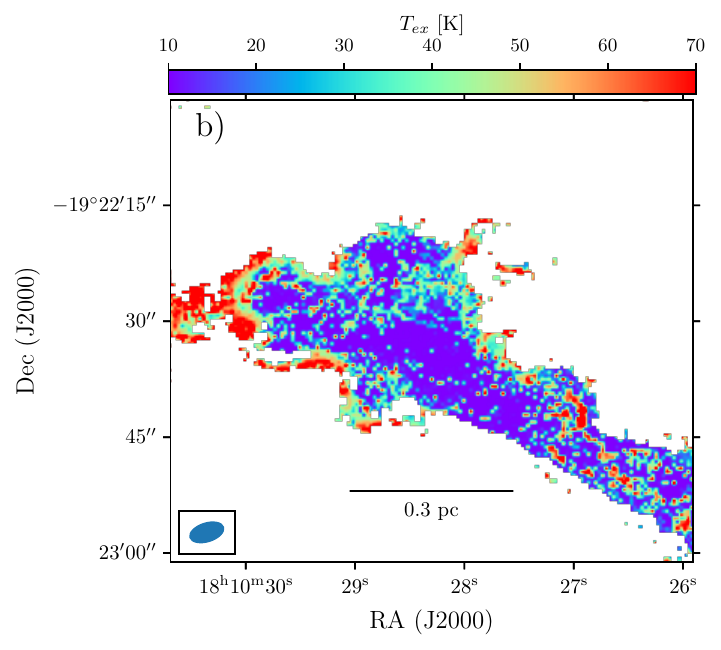}
\end{subfigure}
\caption{a) H$_2$ column density and b) excitation temperature maps of G11P1-HFS. In panel (a), overlaid contour is of N$_{2}$H$^{+}$(1--0) integrated emission at the level of 0.2 Jy beam$^{-1}$ km s$^{-1}$. Star and cross symbols are same as shown in Fig.~\ref{fig1}.
} 
\label{figA2}
\end{figure*}

\begin{figure*}
    \centering
	\includegraphics[width=1\textwidth]{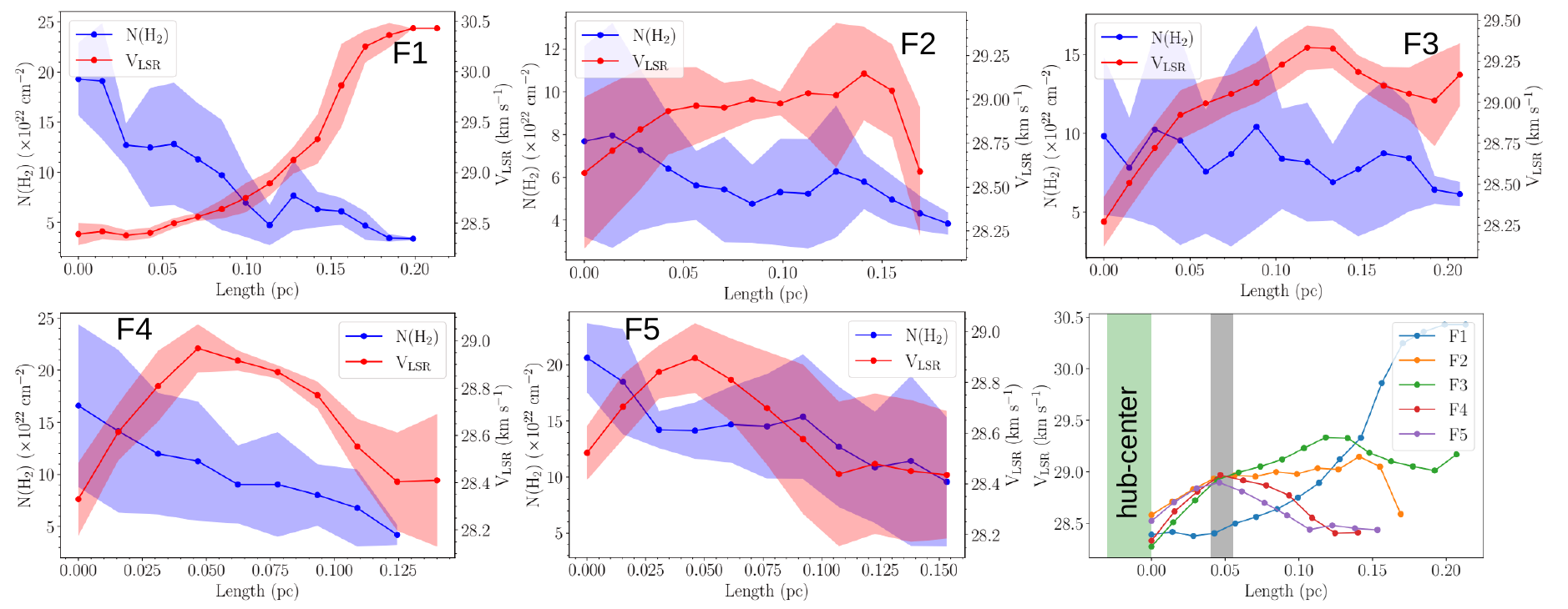}
    \caption {Distribution of $N(\text{H}_{2})$ and $V_{\text{LSR}}$ along the five filaments (F1--F5), from head/hub to tail (see boxes in Fig.~\ref{figA1}c). Shaded regions are same as in Fig.~\ref{fig4}.
    Bottom-right panel shows the $V_{\text{LSR}}$ profiles of all filaments. Shadowed regions mark hub-center and the position range from 0.04 to 0.055 pc.
}
    \label{figA7}
\end{figure*}

\begin{figure*}[htb!]
\begin{subfigure}{0.5\linewidth}
  \centering
\includegraphics[width=1\textwidth]{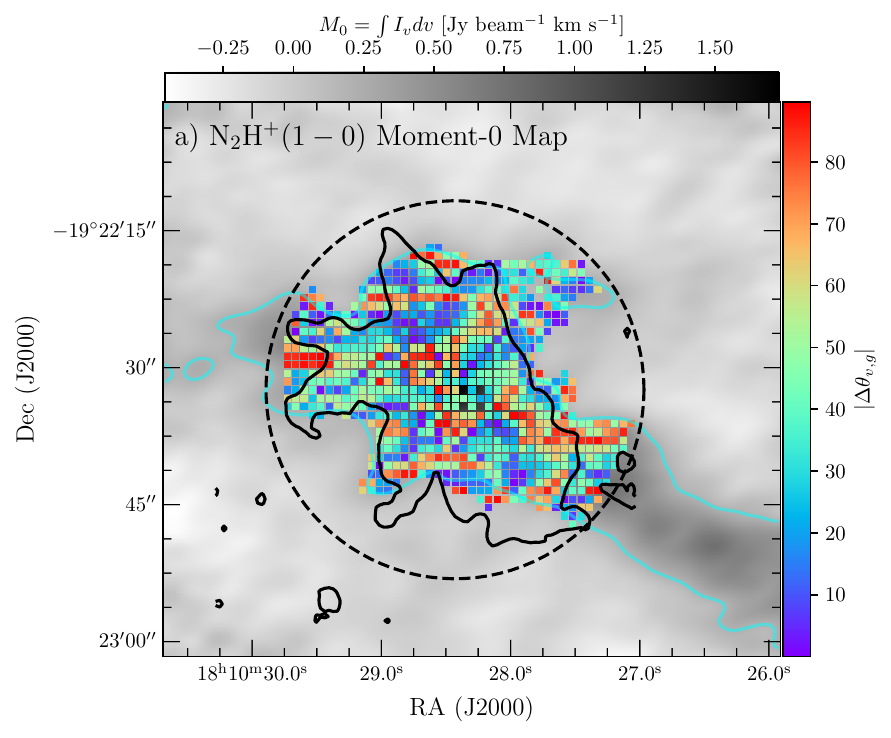}
\end{subfigure}
\begin{subfigure}{0.5\linewidth}
  \centering
\includegraphics[width=1\textwidth]{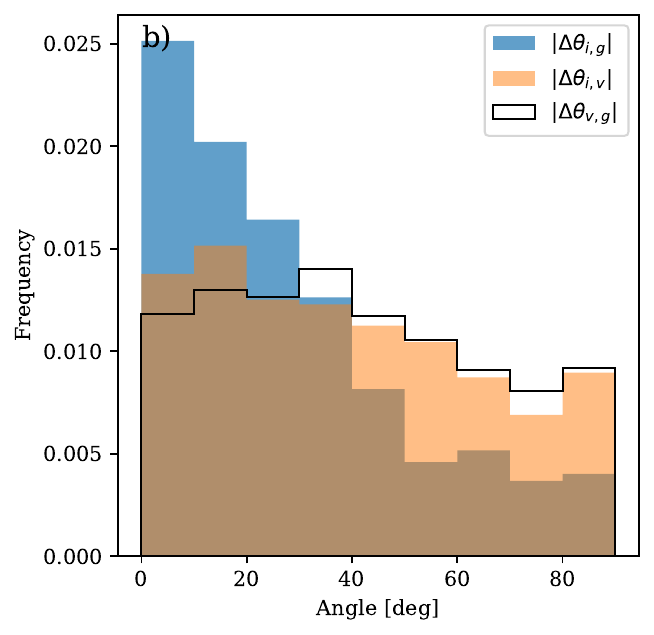}
\end{subfigure}

\caption{a) 
Spatial distribution of relative orientation between {\it g} and {\it v} vectors (in color) overlaid on the N$_{2}$H$^{+}$ moment-0 map.
b) Histogram distribution of relative orientation between {\it g}, {\it v}, and {\it i} vectors toward circular region marked in panel (a). Black and cyan contours are same as the contours shown in Fig.~\ref{figA1}c, and Fig.~\ref{figA2}a, respectively.
} 
\label{figA4}
\end{figure*}

\begin{figure*}
    \centering
	\includegraphics[width=0.5\textwidth]{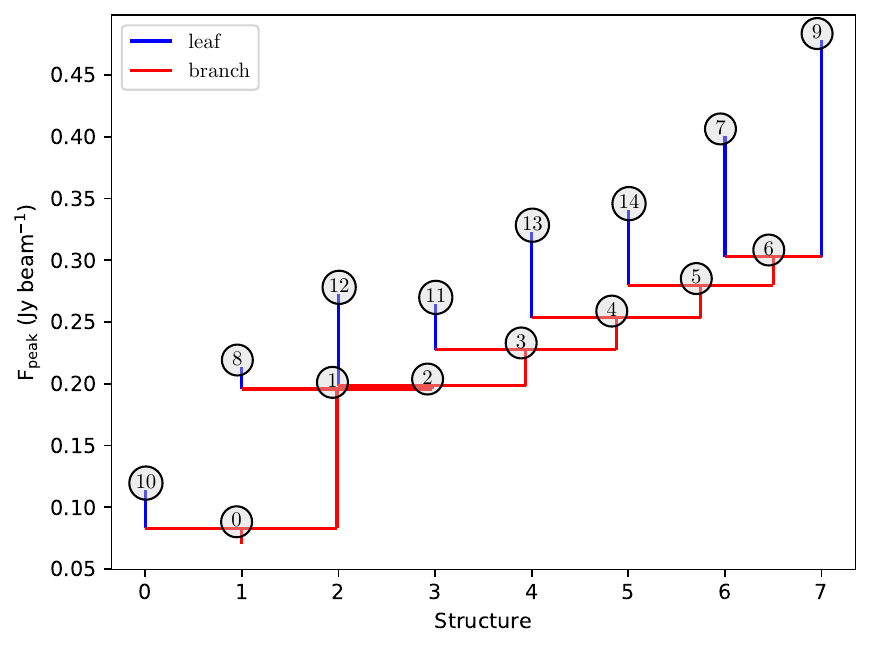}
    \caption {\texttt{Dendrogram} tree of N$_{2}$H$^{+}$(1--0) emission. The leaves and branches are marked (see also Fig.~\ref{figA1}a).
}
    \label{figA3}
\end{figure*}

\begin{table*}[htb!]
\centering
\caption{Physical parameters of \texttt{Dendrogram} leaves for G11P1-HFS}
\label{tab:cores}
\resizebox{\textwidth}{!}{
\begin{tabular}{cccccccccccccccc}
\hline\hline
ID & $RA (J2000)$ & $Dec. (J2000)$ & ${\rm major \times minor}$ axis & PA & $V_{\rm LSR}$ & $\sigma_{\rm obs}$ & $\sigma_{\rm nt}$ & $V_g$ & $T_{\rm ex}$ & $N{\rm (H_{2})}$ & $M$ & $M_{\rm vir}$ & $\alpha_{\rm vir}$ & $\mathcal{M_{\rm 3D}}$ & N \\
  & (hh:mm:sss)  & (dd:mm:sss) &  (${\rm \arcsec  \times \arcsec}$) & (deg) & (km s$^{-1}$) & (km s$^{-1}$) &   (km s$^{-1}$)  &  (km s$^{-1}$ pc$^{-1}$) & (K) & ($\times10^
  {23}$ cm$^{-2}$) & ($M_{\odot}$) & ($M_{\odot}$) & & &  \\
\hline 
L7 & 18:10:28.23 & -19:22:28.29 & 4.67 $\times$ 4.33 & 124.21 & 21.51 & 1.19 & 1.18 & 0.13 & 19.63 & 1.9& 6.5& 38.73& 5.96& 7.84& 1\\
L8 & 18:10:29.52 & -19:22:27.47 & 7.62 $\times$ 2.4 & 172.03 & 20.86 & 0.69 & 0.68 & 0.06 & 19.51 & 0.98& 3.09& 9.27& 3.0& 4.54& 1\\
L9 & 18:10:28.44 & -19:22:33.09 & 11.06 $\times$ 6.62 & 169.5 & 20.29 & 0.83 & 0.83 & 0.06 & 12.6 & 1.8& 22.33& 30.61& 1.37& 6.83& 4\\
L10 & 18:10:30.56 & -19:22:29.99 & 5.18 $\times$ 3.12 & 181.89 & 20.63 & 0.24 & 0.2 & 0.07 & 55.52 & 0.63& 1.54& 1.16& 0.75& 0.78& 1\\
L11 & 18:10:28.86 & -19:22:22.57 & 4.01 $\times$ 3.83 & 118.59 & 20.64 & 0.59 & 0.58 & 0.04 & 25.4 & 1.22& 3.21& 8.35& 2.6& 3.38& 1\\
L12 & 18:10:26.48 & -19:22:49.79 & 7.44 $\times$ 5.14 & 179.3 & 19.56 & 0.66 & 0.65 & 0.04 & 32.13 & 1.31& 8.64& 14.5& 1.68& 3.36& 2\\
L13 & 18:10:27.31 & -19:22:41.55 & 10.72 $\times$ 7.19 & 167.76 & 19.82 & 0.45 & 0.44 & 0.04 & 25.64 & 1.16& 15.33& 9.53& 0.62& 2.55& 4\\
L14 & 18:10:28.72 & -19:22:27.48 & 9.14 $\times$ 2.95 & 191.06 & 20.11 & 1.08 & 1.08 & 0.09 & 23.7 & 1.76& 7.96& 27.84& 3.5& 6.48& 2\\

\hline
\end{tabular}
}
\begin{flushleft}
Note: $V_{\rm LSR}$ and $\sigma_{\rm obs}$ are estimated using the Gaussian fitting of N$_{2}$H$^{+}$(1--0) singlet F$_{\rm 1}$ F = 01--12 at the positions of \texttt{Dendrogram} leaves.
N represents the number of ALMA beams that can fit within the identified structure. $\sigma_{\rm nt}$, $M_{\rm vir}$, $\alpha_{\rm vir}$, and $\mathcal{M}_{\rm 3D}$ denote the non-thermal velocity dispersion, virial mass, virial parameter, and Mach number of the structure, respectively, as estimated following the approach described in \citet{Bhadari2023}.
\end{flushleft}
\end{table*}

\end{appendix}

\end{document}